\documentclass[10pt,conference]{IEEEtran}
\IEEEoverridecommandlockouts
\usepackage{cite}
\usepackage{amsmath,amssymb,amsfonts}
\usepackage{algorithmic}
\usepackage{graphicx}
\usepackage{textcomp}
\usepackage{xcolor}
\usepackage{algorithm2e}
\def\BibTeX{{\rm B\kern-.05em{\sc i\kern-.025em b}\kern-.08em
    T\kern-.1667em\lower.7ex\hbox{E}\kern-.125emX}}
\begin{document}

\title{Scalable Time-Tagged Data Acquisition for Entanglement Distribution in Quantum Networks}

\author{
    \IEEEauthorblockN{
        Abderrahim Amlou\IEEEauthorrefmark{1}\IEEEauthorrefmark{2},
        Thomas Gerrits\IEEEauthorrefmark{1},
        Anouar Rahmouni\IEEEauthorrefmark{1},
        Amar Abane\IEEEauthorrefmark{1},\\
        Mheni Merzouki\IEEEauthorrefmark{1},
        Ya-Shian Li-Baboud\IEEEauthorrefmark{1},
        Ahmed Lbath\IEEEauthorrefmark{2},
        Abdella Battou\IEEEauthorrefmark{1},
        Oliver Slattery\IEEEauthorrefmark{1}
    }
    \IEEEauthorblockA{
        \IEEEauthorrefmark{1}National Institute of Standards and Technology (NIST), Gaithersburg, MD, USA\\
        \IEEEauthorrefmark{2}Univ. Grenoble Alpes, Grenoble, France\\
        Corresponding author: Abderrahim Amlou (abderrahim.amlou@nist.gov)
    }
}

\maketitle

\begin{abstract}
% In distributed quantum applications such as entanglement distribution and photon-based communication, precise time synchronization and efficient time-tagged data handling are critical. Traditional time tagging systems face significant limitations, including timestamp overflows, synchronization drift, and inefficient data storage, which hinder scalability and long-term operation. 

% This paper presents a modular Time Tagging (TT) agent designed to optimize data acquisition in such systems. The agent operates using a 1 pulse-per-second (PPS) reference signal from White Rabbit (WR) devices to ensure network-wide synchronization, while applying real-time calibration, overflow mitigation, and compression to the collected data.

% We demonstrate the effectiveness of the system through a live entanglement distribution experiment between two laboratories, where the TT agents streamed data to a coincidence analysis agent in real time. The system maintained precise synchronization, enabling the detection of approximately 25,000 entangled photon pairs per second within a 10 ns coincidence window. In parallel, data preparation and compression reduced storage costs from 14.32 to 3.80 bytes per tag. These results confirm that the proposed approach supports scalable, efficient, and accurate time-tagged data acquisition across distributed quantum networks.

In distributed quantum applications such as entanglement distribution, precise time synchronization and efficient time-tagged data handling are essential. Traditional systems often suffer from overflow, synchronization drift, and storage inefficiencies.
We propose a modular Time Tagging (TT) agent that uses a 1 pulse per second (PPS) signal from White Rabbit (WR) devices to achieve network-wide synchronization, while applying real-time calibration, overflow mitigation, and compression. A live two-lab entanglement distribution experiment validated the system’s performance, achieving synchronized coincidence detection at 25,000 counts/sec within a 10 ns window. Data optimization reduced the storage and classical communication bandwidth requirement from 14.32 to 3.80 bytes per tag. The results demonstrate a scalable, accurate, and efficient solution for distributed quantum networks.
\end{abstract}


\section{Introduction}

Quantum networks are rapidly emerging as a key technology for enabling entanglement-based applications such as distributed quantum computation and quantum key distribution (QKD). At the core of these networks are single photons, which serve as quantum information carriers between remote nodes \cite{lim2005quantum}. Ensuring that these photons are detected and correlated precisely across distributed locations is critical for the functionality of such systems.

Achieving this, high-resolution time tagging is essential. Time taggers are devices that record the exact arrival times of single-photon detection events with sub-nanosecond precision. These time-stamped events are essential for verifying entanglement, performing coincidence analysis, and supporting synchronization in quantum protocols \cite{katz2018coincidence}.

Some of the major challenges in time tagging systems relate to ensuring precise synchronization between distributed time taggers, either due to the lack of a common synchronization system or to cumulative timing inaccuracies introduced by internal clock drift \cite{synchronisation}. Additionally, these systems can only operate reliably over finite periods before the risk of overflow becomes significant, limiting their usability for long-term applications where uninterrupted, high-precision data acquisition is essential. Another challenge lies in handling high data volumes. The continuous streaming of time-tagged data, often accompanied by redundant metadata, places significant demands on storage and network resources, particularly as data from multiple channels accumulates over time. These issues lead to storage constraints and complicate data access and retrieval, ultimately limiting the scalability of time-tagging systems in large networks. \cite{swabianManual}

To address these challenges, we propose a novel Time Tagging (TT) agent that enhances data acquisition by performing real-time synchronization, calibration, overflow mitigation, and compression directly at the source of the data. As a result, the time-tagged data is fully prepared and ready for analysis, enabling scalable and efficient operation in distributed quantum systems.

The remainder of this paper is organized as follows. Section~\ref{sec:background} provides background on time-tagging systems and outlines the key challenges associated with synchronization, overflow, and data handling in distributed quantum applications. Section~\ref{sec:approach} presents our proposed system architecture and methodology, including real-time synchronization, calibration, overflow mitigation, and compression. Section~\ref{sec:evaluation} describes the experimental setup and evaluates the performance of our system through live entanglement distribution and coincidence analysis. Section~\ref{sec:discussion} discusses the implications of the results. Finally, Section~\ref{sec:conclusion} concludes the paper.

\label{sec:approach}
\label{sec:evaluation}
\label{sec:discussion}
\label{sec:future_work}
\label{sec:conclusion}
\section{Background} \label{sec:background}

In our previous setup, time taggers were used alongside single-photon detectors to record the precise arrival times of detected photons. A time tagger is an electronic device that generates high-resolution timestamps upon receiving input signals, typically from detectors \cite{swabianManual}. In quantum experiments, especially those involving entangled photon pairs, time taggers are essential for identifying correlated detection events across remote stations \cite{photon_detect_with_tt}.

\begin{figure}[ht]
    \centering
    \includegraphics[width=\columnwidth]{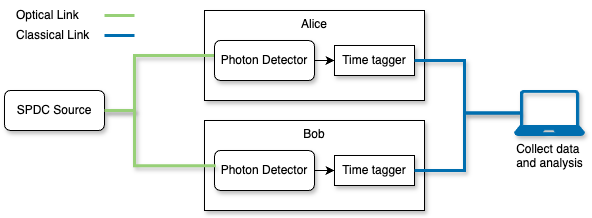}
    \caption{The SPDC source generates entangled photon pairs sent to Alice and Bob, where each detector is connected to a time tagger that records photon arrivals for later analysis.}
    \label{fig:spdc_setup}
\end{figure}

The system was applied to an entanglement distribution experiment between two nodes (e.g., Alice and Bob), using a spontaneous parametric down-conversion (SPDC) source to produce entangled photon pairs \cite{rahmouni2024entangled}, as illustrated in Figure~\ref{fig:spdc_setup}. In this context, synchronized and accurate timing information is crucial for performing coincidence analysis as a part of verifying the presence of entanglement between the two parties.

Despite its effectiveness in time-stamping photon events, the previous system faced several limitations that hindered its scalability and precision for long-term operation.

\subsection{Synchronization Challenges}
Achieving precise synchronization between distributed time taggers is a fundamental challenge in time-sensitive applications. In the absence of a shared, highly accurate timing reference, timestamps generated by different time taggers may drift relative to each other, making it difficult to align and analyze events recorded at separate locations. Clock drift, variations in local oscillators, and differences in data transfer protocols further amplify these discrepancies, and timing inaccuracies. This synchronization challenge poses significant obstacles to accurately correlating data from multiple and remote time taggers.

\subsection{Overflow Issues}

Many different time taggers and data transfer protocols exist today~\cite{swabianSeries,picoharp330,rossetta2021brighteyes}. These devices typically represent timestamps using a finite number of bits, which limits the maximum duration over which events can be recorded. For instance, a 64-bit timestamp representation~\cite{swabianManual} allows reliable data acquisition for approximately 106 days with 1 ps resolution \cite{swabianManual} before reaching its maximum value and risking overflow. While the device may wrap around internally, in distributed experiments where multiple systems must remain synchronized, such overflows can cause inconsistencies or data misalignment. Without mitigation, this requires manually resetting each connected node to prevent system-wide data loss or drift, as illustrated in the heatmap in Figure~\ref{fig:heatmap}.

\subsection{Clock Drift and Calibration Requirements}

Another critical issue in time-tagged systems is the gradual drift of the internal clock within each time tagger. Even when two devices are synchronized to a common reference (e.g., 1 pulse per second (PPS) signal), slight discrepancies in their local oscillators can lead to time inaccuracies that accumulate over long durations~\cite{ho2009clock, enzer2016drifts}. These drifts affect the precision of timestamped events, making it difficult to maintain consistent alignment between distributed nodes over time.

In high-resolution applications such as entanglement verification, even small drift-induced offsets can significantly reduce the quality of coincidence analysis~\cite{enzer2016drifts}. As a result, regular calibration is required to correct for timing inconsistencies and ensure that time tags remain accurate and comparable across all devices in the network.

The need for such calibration becomes even more critical in long-term or high-rate experiments, where the timestamp precision must be preserved over millions of detection events.

\subsection{Data Handling Inefficiencies}

As detection events accumulate, continuous data streaming places a heavy demand on both storage and network resources. Without efficient data handling strategies, excessive data generation can lead to storage bottlenecks, increased retrieval times, and network congestion, ultimately slowing down the analysis.
Additionally, some data handling systems introduce unnecessary metadata or redundant information, further inflating storage requirements and making access to critical data less efficient. The problem is compounded when using multiple input channels on a time tagger, where each channel receives events from a separate detector. As the number of channels increases, so does redundancy and data volume, making it harder to manage and process the data efficiently. Without effective compression or optimized structuring methods, data handling becomes impractical for long-term, high-throughput data acquisition, particularly in distributed setups that demand scalable and timely data processing.

\section{Proposed Approach} \label{sec:approach}

In this section, we present the approach and system design for optimizing time-tagging systems in distributed, time-sensitive applications. The proposed system is specifically designed for distributed quantum networks and similar applications that demand high-precision time-tagged data acquisition. It is also agnostic to the type of time tagger used, allowing integration with various hardware platforms.

The discussion begins with an overview of the system architecture, highlighting its components and how they collectively enable precise synchronization and scalable data acquisition. We then describe the proposed approach, focusing on four key aspects: \textbf{synchronization, calibration, overflow mitigation, and data compression}.

\subsection{System Architecture}
\begin{figure*}[ht]
    \centering
    \includegraphics[width=\textwidth]{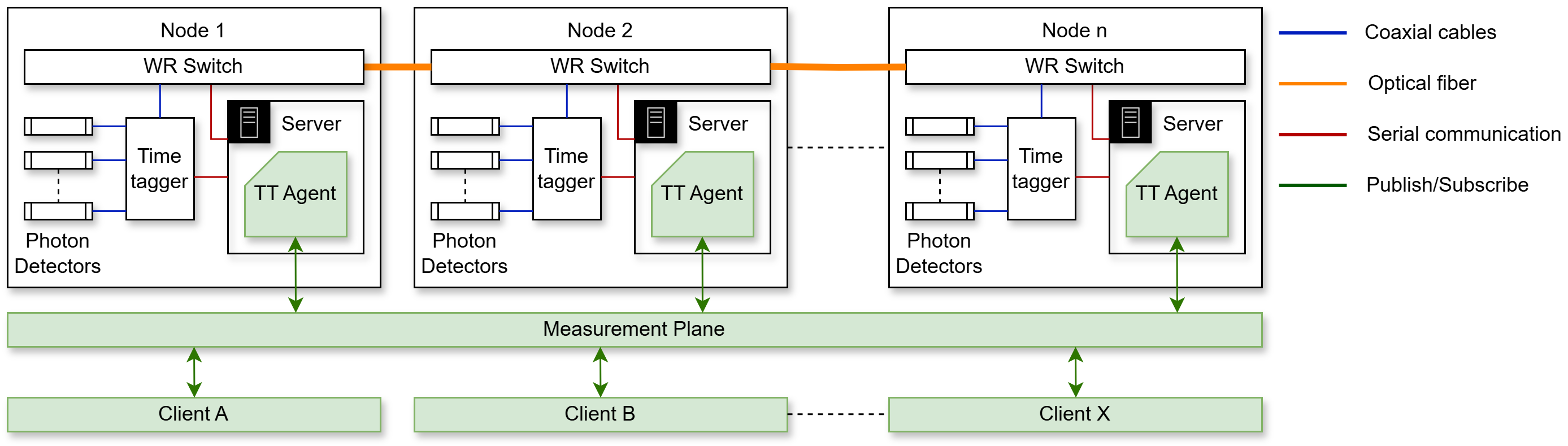}
    \caption{System Architecture of the Distributed Time Tagging Setup.}
    \label{fig:archi}
\end{figure*}

To support scalable, high-precision time-tagged data acquisition across distributed quantum nodes, the proposed system follows a modular architecture that enables local processing, network-wide synchronization, and real-time client interaction. Figure~\ref{fig:archi} illustrates the structure of each node in the system.

Each node includes a \textbf{time tagger} connected to a photon detector and the White Rabbit (WR) switch, enabling high-precision timestamping and synchronization via 1 PPS pulses. A server at the node runs the \textbf{TT agent}, which pulls and processes data from these components.

The TT agent at each node communicates through the measurement plane \cite{abane2023data}, which serves as an interface to manage measurements and collect data. 
According to the measurement plane, each TT agent operates automatically to:
\begin{itemize}
    \item Advertise its time tags measurement capability to clients.
    \item Receive client requests specifying measurement parameters, such as time scope and specific channels.
    \item Process and prepare the requested time tags data.
    \item Send the processed data back to the client.
\end{itemize}

\subsection{Key Components and Functionalities}
The proposed approach introduces a Time Tagger (TT) agent to optimize data acquisition from the time tagger, focusing on synchronization, calibration, overflow mitigation, and data efficiency. This section details the main components of the new approach, which relies on using a 1 pulse-per-second (PPS) reference signal from a White Rabbit device for real-time data processing.\\

This section details the main components of the new approach.

\subsubsection{Synchronization}\hfill\\
The core of the proposed approach is based on the use of a 1 PPS reference signal provided by a White Rabbit device (or any other high-accuracy technology) that is synchronized throughout the network. This signal acts as a common timing reference for all distributed nodes, enabling the TT agent to organize and align the sequences of time tags with an absolute time reference for each second.

As illustrated in the example in Figure \ref{fig:dataprocessing}, upon receiving the 1 PPS pulse, the TT agent marks this timestamp as the time reference for the upcoming sequence of time tags, which are collected until the arrival of the next 1 PPS signal. The agent then assigns the entire sub-sequence of time tags to the absolute time of the corresponding first 1 WR pulse in that interval. And this absolute time is obtained directly from the White Rabbit device. This approach provides a consistent and synchronized organization of time-tagged data across the network.\\

\subsubsection{Overflow Mitigation}\hfill\\
The proposed approach also includes an overflow mitigation mechanism to address the limitations of time tag accumulation over long periods. To prevent overflow, the TT agent resets the reference point for each sequence of time tags based on the synchronized 1 PPS signal. Using the timestamp of each new 1 PPS pulse as a reference, the TT agent defines each time tag within an interval as a relative value rather than an absolute timestamp. This relative timing is achieved by subtracting the timestamp of the current 1 PPS pulse from each time tag in the sequence:
\[
\text{TT}_{\text{relative}} = \text{TT}_{\text{cal}} - \text{PPS}_{\text{cur}}
\]

where \( \text{PPS}_{\text{cur}} \) is the timestamp of the latest 1 PPS pulse, and \( \text{TT}_{\text{cal}} \) is the calibrated time tag value.

\begin{figure}[ht]
    \centering
    \includegraphics[width=\columnwidth]{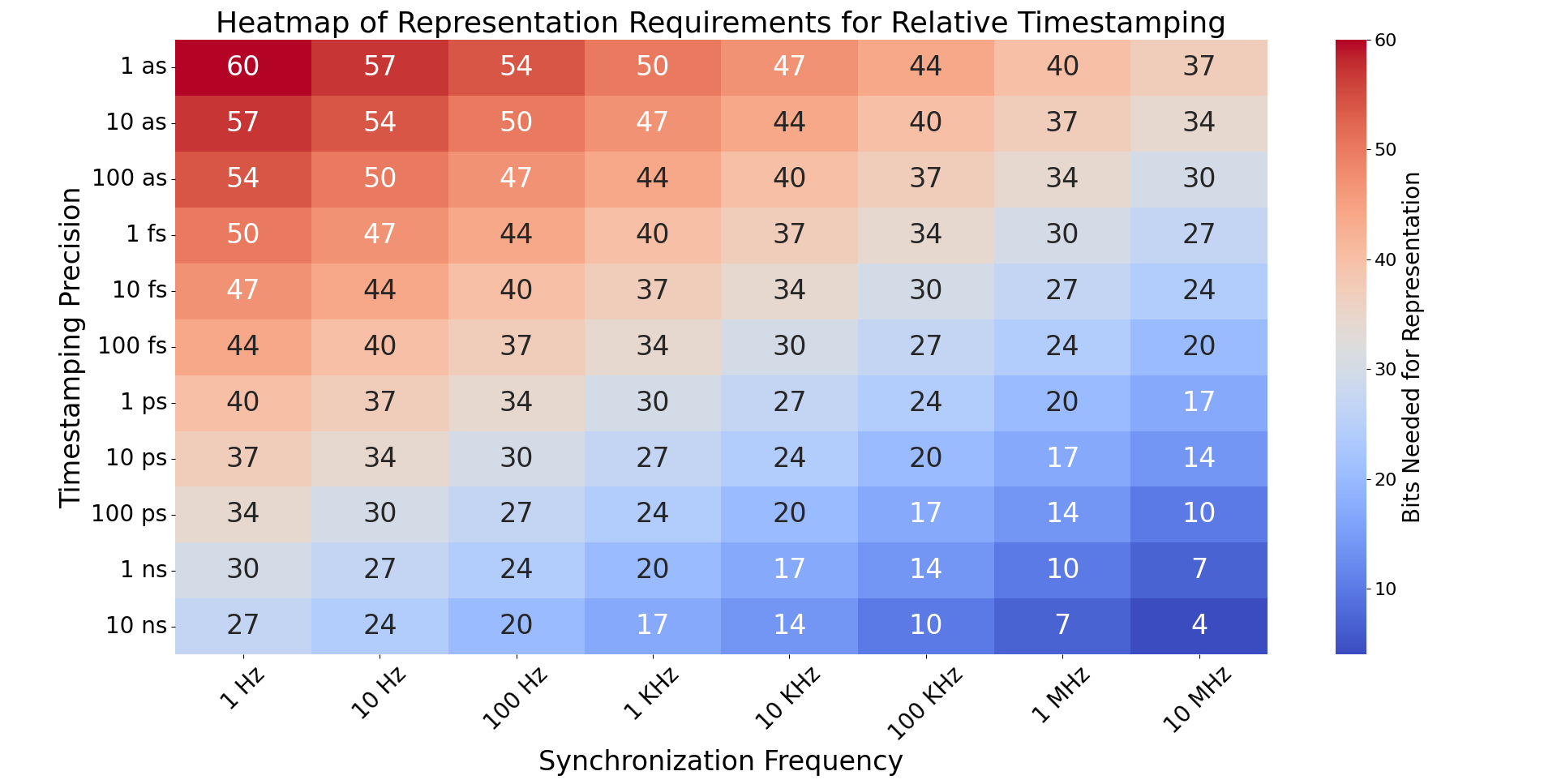}
    \caption{Heatmap of representation requirements for relative time stamping assuming minimum resolution of 1 ps.}
    \label{fig:heatmap}
\end{figure}

By storing time tags as relative values within each interval, the system avoids the accumulation of large absolute timestamps, effectively preventing overflow. Each sub-sequence of time tags is associated and identified with the 1 PPS event's WR absolute time that serves as its reference, ensuring that the time tags remain within a manageable range. Since these relative timestamps are always limited to a 1-second interval, their bit representation is significantly reduced. For example, at a resolution of 1 picosecond (ps), only 40 bits are required for timestamp representation by the time tagger (see Figure \ref{fig:heatmap}).\\

This allows the system to operate without interruption or risk of overflow, providing consistent and reliable time-tags across all nodes in the network.
\\

The efficiency of relative timestamping depends on the synchronization frequency and time resolution. Figure \ref{fig:heatmap} illustrates how resolution and interval length affect bit-width requirements.

\subsubsection{Calibration} \hfill\\
In addition to synchronization, the proposed approach includes a real-time calibration process to enhance the accuracy of each time tag within its designated interval. Calibration is necessary due to drift in the internal clock of the time tagger, which can introduce timing inconsistencies over extended periods. To correct for this drift, the TT agent applies a calibration factor to each time tag, adjusting it based on the timing consistency of the 1 PPS signals.

For each sequence of time tags, the calibration factor \( \text{CF} \) is calculated as:

\[
\text{CF} = \frac{1}{\text{PPS}_{\text{cur}} - \text{PPS}_{\text{old}}}
\]

where \( \text{PPS}_{\text{cur}} \) and \( \text{PPS}_{\text{old}} \) represent the timestamps of the most recent and the preceding 1 PPS pulses, respectively. Each time tag is calibrated using the last two consecutive 1 PPS timestamps to account for small variations in the PPS interval. This calibration factor compensates for any slight deviations in the timing of consecutive PPS signals.

Each time tag \( \text{TT}_{\text{relative}} \) is then calibrated by applying this factor:

\[
\text{TT}_{\text{cal}} = \text{CF} \times \text{TT}_{\text{relative}}
\]

This method ensures that each time tag is adjusted according to the precise interval between consecutive PPS signals, enhancing timing accuracy across all distributed nodes. Thus, the system provides consistently accurate time-tagged data that is ready for analysis without additional post-processing.\\

\subsubsection{Data Preparation and Compression}\hfill\\
The proposed methodology incorporates data preparation and compression to enhance efficiency. Without these optimizations, raw data from time taggers may contain redundant metadata that does not need to be transmitted. This excess data increases storage requirements and unnecessarily overloads the network, ultimately reducing overall system performance.

In contrast, the TT agent selectively filters the data before processing, retaining only the essential time tags needed for synchronization and analysis. By discarding redundant metadata from the data stream, the TT agent ensures that only meaningful, calibrated, and relative time tags are retained. This reduction in data volume significantly enhances the effectiveness of the compression process, allowing it to achieve higher compression ratios with less computational overhead.

The TT agent employs Blosc, a high-performance, multi-threaded compression library, to compress this filtered data before it is streamed or stored. Blosc’s in-memory compression and decompression capabilities allow for rapid data handling, making it particularly suitable for time-sensitive applications. By applying Blosc compression to the already-filtered data, the TT agent reduces storage and network load. In practice, this achieves over 70\% compression rate on average (see Section~\ref{sec:evaluation}), significantly reducing storage and network load to support scalable, long-term acquisition in distributed setups.
\subsection{Prototype Implementation and Workflow:}
The TT agent is implemented in Python and runs as a background service on each node's server. It communicates directly with the local time tagger and White Rabbit (WR) device, operating in real time to synchronize, calibrate, compress, and organize time-tagged data.

\begin{figure}[ht]
    \centering
    \includegraphics[width=\columnwidth]{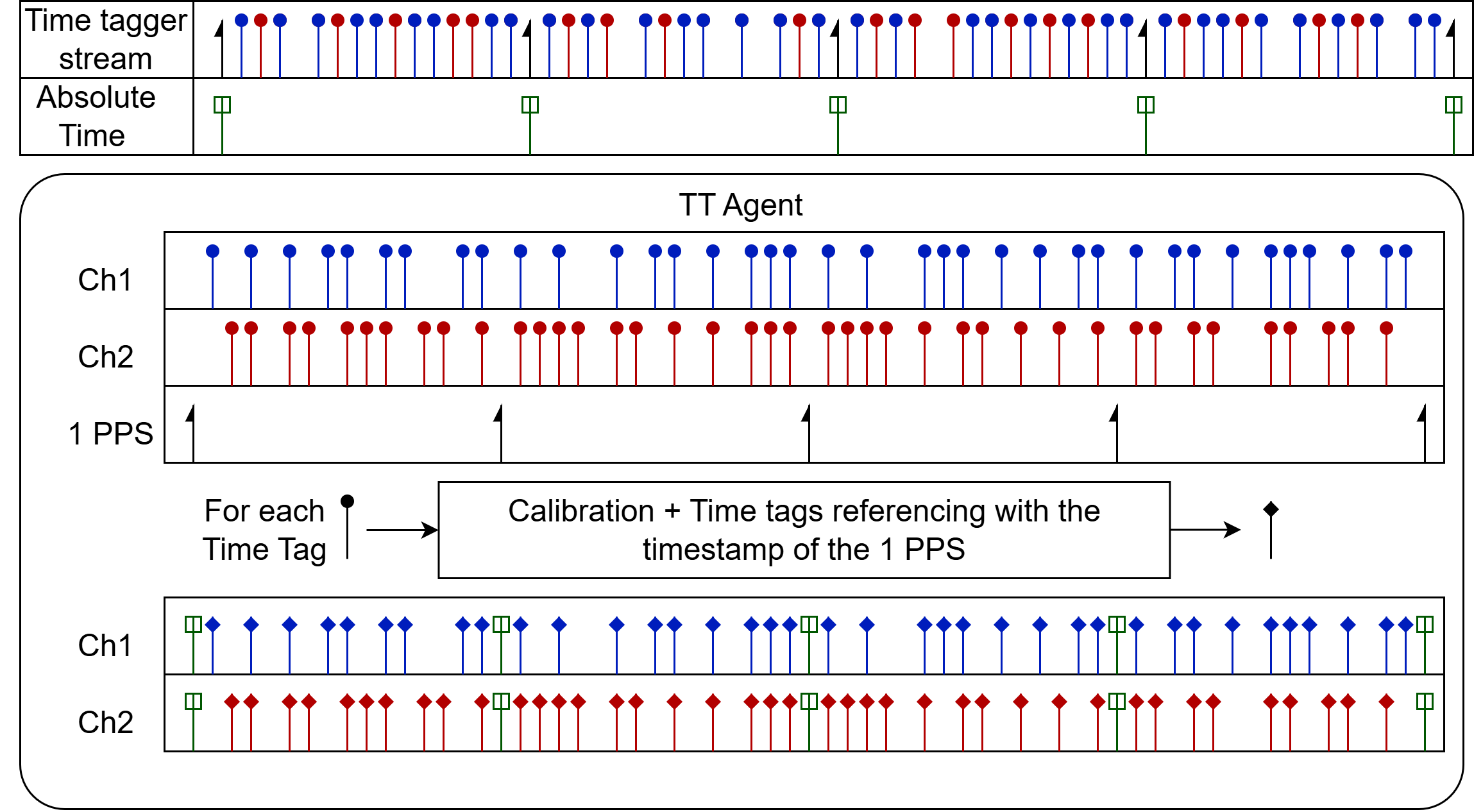}
    \caption{Real Time Data Processing Workflow.}
    \label{fig:dataprocessing}
\end{figure}

As shown in Figure~\ref{fig:dataprocessing}, the agent continuously pulls timestamped events from the time tagger, typically receiving millions of events per second across multiple channels. Upon receiving each 1 PPS signal from the WR device, it queries the WR for the absolute time and processes the corresponding 1-second data block as follows: Timestamps are referenced relative to the 1 PPS signal to ensure consistency and prevent overflow; calibration is applied to correct internal clock drift using the interval between consecutive PPS pulses; the resulting time tags are then associated with the absolute time of the PPS; data is organized by second and channel and compressed using the Blosc library; finally, the prepared data is streamed to the client via the measurement plane. This per-second processing pipeline ensures synchronized, calibrated, and compressed data is available in real time and can be easily identified and retrieved afterward.

% By following this workflow, the TT agent ensures that time-tagged data is accurately prepared, synchronized, and efficiently delivered for analysis.

\section{Evaluation and Results} \label{sec:evaluation}
In this section, we evaluate the performance of the proposed approach in a real-world quantum networking scenario. We conducted a live entanglement distribution experiment between two laboratories. The goal is to verify whether the system could maintain precise synchronization and enable real-time coincidence analysis across distributed locations—ultimately confirming the successful generation of entanglement.

\begin{figure*}[ht]
    \centering
    \includegraphics[width=0.9\textwidth]{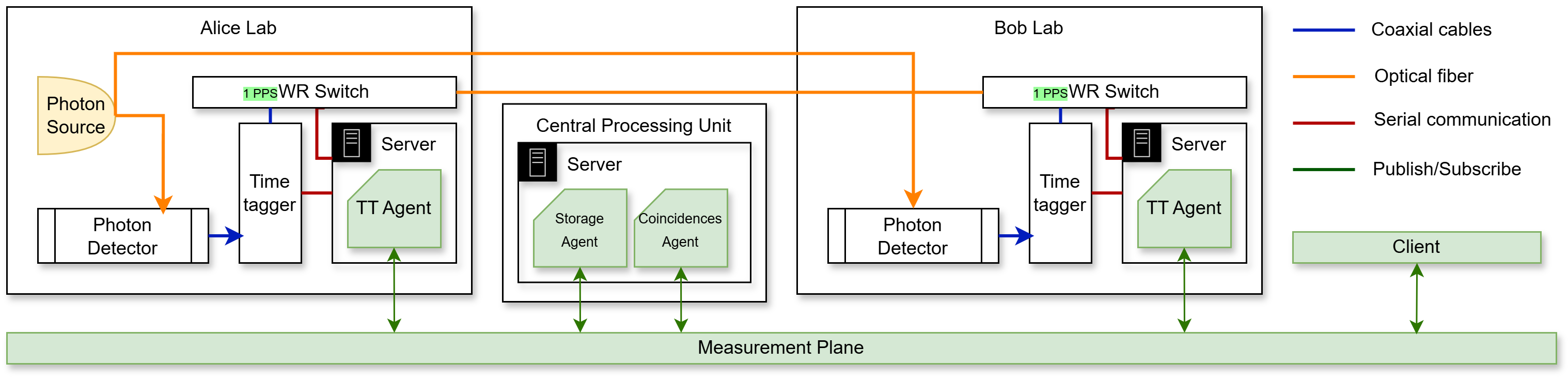}
    \caption{Experimental Setup.}
    \label{fig:experiment}
\end{figure*}
\subsection{Experimental Setup}
As illustrated in Figure \ref{fig:experiment}, the experiment involves two laboratory setups, each equipped with with a\textbf{ White Rabbit (WR)} switch for timing, a \textbf{time tagger} connected to a \textbf{single-photon detector}, and a \textbf{server} running the TT agent.

A spontaneous parametric down-conversion (SPDC) source was used to generate pairs of entangled photons, referred to as the signal and idler. The signal photon was sent to Alice's Lab, and the idler photon to Bob's Lab, using single-mode optical fiber.

The TT agents at both labs received the 1 PPS signal as a reference clock and continuously pulled time tags from the local time taggers. Time tags were processed in real time, calibrated, referenced to the WR absolute time, and compressed before being passed to a central coincidence analysis agent.

\subsection{Live Coincidence Analysis}

To evaluate entanglement between the two labs, we used a dedicated coincidences agent developed as part of the measurement plane. This agent communicates directly with the TT agents at both locations to initiate measurements and stream synchronized time-tagged data in real time.

Once the measurement is started, the coincidences agent receives processed time tags from the TT agents in real time. It then performs live coincidence analysis, returning the coincidences results between Alice and Bob to the client.

The analysis compared time-tagged detection events from both labs in real time. By aligning the data using the absolute times of synchronized 1 PPS references, we could accurately match detection events that occurred simultaneously at both locations.

The key indicator of success was the appearance of a coincidence peak in the time difference histogram between Alice's Lab and Bob's Lab events. This peak represents the correlated detection of entangled photons at the detectors in both labs.

\begin{figure}[ht]
    \centering
        \includegraphics[width=\columnwidth]{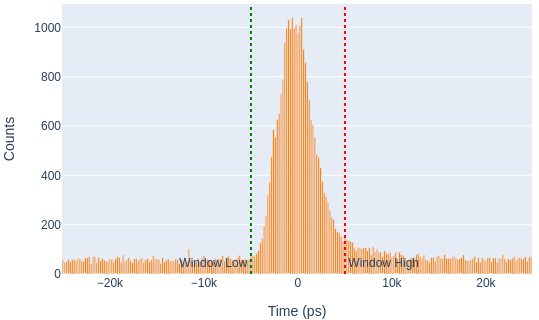}
    \caption{Live coincidence histogram (1-second time frame) between Alice and Bob}
    \label{fig:1s_peak}
\end{figure}

\begin{figure}[ht]
    \centering
        \includegraphics[width=\columnwidth]{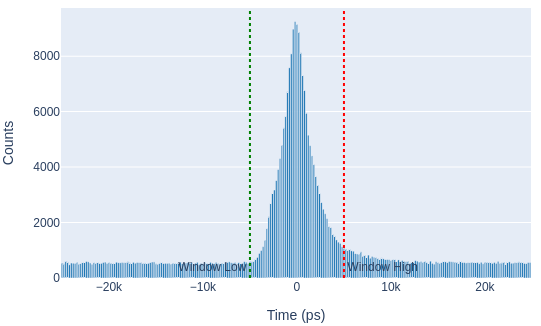}
    \caption{Accumulated coincidence histogram over multiple intervals.}
    \label{fig:accumulative_peak}
\end{figure}

The coincidence histogram initially revealed a clear and narrow peak at approximately $7 \mu s$, which corresponds to the expected time delay between the two labs due to optical fiber length differences and hardware-induced offsets. After estimating this delay through preliminary coincidence analysis, we applied a channel delay compensation directly in the time tagger configuration in Alice's lab to align the detection events. Following this adjustment, as shown in Figure \ref{fig:1s_peak} and \ref{fig:accumulative_peak}, the coincidence peak shifted to 0 ps, confirming that both time taggers are now synchronized at the event level. 

Then we performed real-time coincidence analysis. Figure \ref{fig:1s_peak} shows a snapshot of the live coincidences histogram, capturing events within a one-second interval. And to track stability over time, we also accumulated the coincidences from each live result. Figure \ref{fig:accumulative_peak} presents the accumulated coincidences histogram, where the peak becomes more pronounced as new one-second data chunks are continuously added.


\begin{figure}[ht]
    \centering
        \includegraphics[width=\columnwidth]{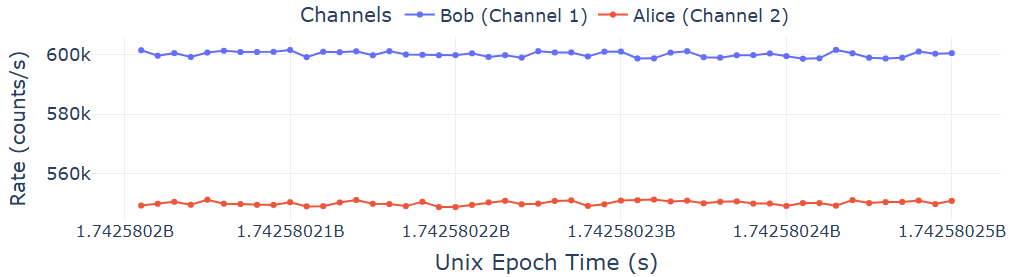}
        \vspace{0.5em} % optional space between the two images

        \includegraphics[width=\columnwidth]{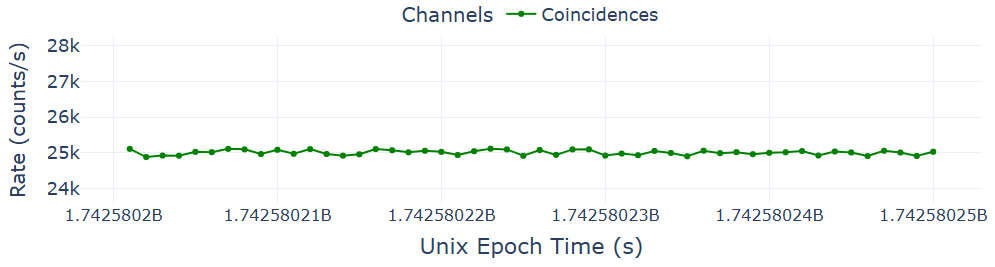}
    \caption{Live count rates: Bob (Ch1), Alice (Ch2), and coincidence rate.}
    \label{fig:ch_rates}
\end{figure}

\begin{itemize}
    \item Channel 1 (Bob): $\approx 600{,}000$  counts per second

    \item Channel 2 (Alice):  $\approx 550{,}000$ counts per second

    \item Coincidence Rate:  $\approx 25{,}000$ coincidences per second.
\end{itemize}

In addition to the histogram analysis, we continuously monitored the detection rates of each channel and the coincidence rate. Figure \ref{fig:ch_rates} shows the recorded photon's detection rates and coincidence rate.
With a 10 ns coincidence window, we were able to identify approximately 25,000 correlated photon pairs per second, which corresponds approximately to $\approx 4$ \% of the detection rate. This level of coincidence efficiency is consistent with what is typically expected in fiber-based entanglement distribution experiments and confirms the high quality of both the entanglement source and the timing synchronization.

\subsection{Data Preparation and Compression Efficiency}
To evaluate the effect of the TT agent’s data preparation and compression steps, we compared the storage cost per time tag in two scenarios: with and without preparation/compression.
Normally, without any data preparation or compression, each time tag costs approximately 14.32 bytes, including a large amount of metadata. After applying the TT agent’s filtering and compression pipeline, this drops to 3.80 bytes per tag, resulting in a 73.5\% reduction in storage size. This significant improvement enables scalable, long-term operation in bandwidth and storage-constrained environments.

\section{Discussion} \label{sec:discussion}
The proposed Time tagger (TT) agent successfully addresses key challenges in distributed time-tagged data acquisition, including synchronization, overflow handling, and efficient storage. Through the integration of a synchronized 1 PPS reference from White Rabbit devices and a per-second real-time processing pipeline, the TT agent enables precise alignment of time-tagged events across remote nodes.

The live entanglement distribution experiment between two labs confirms the system’s ability to operate in a realistic quantum networking environment. The coincidence peak observed at 0~ps, after delay compensation, validates synchronization at the event level. Additionally, the consistent coincidence rate of approximately 25,000 counts per second across an extended period demonstrates the system's stability and readiness for real-time quantum experiments.

One limitation observed in our evaluation is the slight skew and broadening of the coincidence peak in Figure~\ref{fig:1s_peak}. This effect is likely due to drifts of the time tagger's internal clock within the 1-second pulse interval. The current system uses a 1 PPS signal to calibrate each second of data, which aligns timing across devices but does not compensate for clock drift occurring within a 1-second interval. Since our evaluation setup uses a Swabian Time Tagger 20 with a free-running oscillator, such drift can introduce timing uncertainty. To mitigate this limitation, future implementations could discipline the time tagger clock using a 10~MHz reference output from the White Rabbit switch, improving stability and reducing intra-second drift effects.

The storage evaluation also confirms the benefits of integrating data preparation and compression into the acquisition pipeline. The 73.5\% improvement is significant, especially for large-scale and long-duration measurements. This enables the deployment of time-tagging systems in environments with limited storage or bandwidth resources.

Moreover, the modular integration of the TT agent within a measurement plane architecture allows for scalable expansion and real-time monitoring, without requiring centralized post-processing. This decentralized design supports dynamic experimentation, making it adaptable to various quantum network scenarios.

\section{Conclusion} \label{sec:conclusion}
This paper presents a complete approach for improving time-tagged data acquisition in distributed, time-sensitive quantum applications. By combining precise synchronization using a 1 PPS signal, real-time calibration, overflow mitigation, and efficient data compression, the proposed TT agent offers a scalable and reliable solution for long-term operation.

The system was validated through an experimental setup across two laboratories, confirming its practical applicability in real quantum network scenarios. By integrating with a measurement plane and supporting automated, distributed operation, the solution provides a solid foundation for future quantum network deployments.

The integration of the TT and coincidence agents within a measurement plane framework allows researchers to automate and scale measurements across distributed networks. This makes the system suitable not only for current experiments, but also as a foundation for more advanced quantum networking protocols. Future developments will further extend its capabilities toward multi-node setups, higher-rate sources, and integration with complete quantum communication stacks.

\section{Disclaimer} \label{sec:DISCLAIMER}
Any mention of commercial products or reference to commercial organizations is for information only; it does not imply
recommendation or endorsement by NIST, nor does it imply
that the products mentioned are necessarily the best available
for the purpose.

% Generated by IEEEtran.bst, version: 1.14 (2015/08/26)


\begin{thebibliography}{10}
\providecommand{\url}[1]{#1}
\csname url@samestyle\endcsname
\providecommand{\newblock}{\relax}
\providecommand{\bibinfo}[2]{#2}
\providecommand{\BIBentrySTDinterwordspacing}{\spaceskip=0pt\relax}
\providecommand{\BIBentryALTinterwordstretchfactor}{4}
\providecommand{\BIBentryALTinterwordspacing}{\spaceskip=\fontdimen2\font plus
\BIBentryALTinterwordstretchfactor\fontdimen3\font minus \fontdimen4\font\relax}
\providecommand{\BIBforeignlanguage}[2]{{%
\expandafter\ifx\csname l@#1\endcsname\relax
\typeout{** WARNING: IEEEtran.bst: No hyphenation pattern has been}%
\typeout{** loaded for the language `#1'. Using the pattern for}%
\typeout{** the default language instead.}%
\else
\language=\csname l@#1\endcsname
\fi
#2}}
\providecommand{\BIBdecl}{\relax}
\BIBdecl

\bibitem{lim2005quantum}
Y.~L. Lim, ``Quantum information processing with single photons,'' \emph{arXiv preprint quant-ph/0509168}, 2005.

\bibitem{katz2018coincidence}
E.~J. Katz, N.~C. Wilson, I.~R. Nemitz, S.~A. Tedder, B.~E. Vyhnalek, B.~M. Floyd, T.~D. Roberts, P.~Pandit, S.~Baugher, R.~P. Tokars \emph{et~al.}, ``Coincidence studies of entangled photon pairs using nanowire detection and high-resolution time tagging for qkd application,'' in \emph{Broadband Access Communication Technologies XII}, vol. 10559.\hskip 1em plus 0.5em minus 0.4em\relax SPIE, 2018, pp. 14--24.

\bibitem{synchronisation}
S.~S. Nande, M.~Paul, S.~Senk, M.~Ulbricht, R.~Bassoli, F.~H. Fitzek, and H.~Boche, ``Quantum enhanced time synchronisation for communication network,'' \emph{Computer Networks}, vol. 229, p. 109772, 2023.

\bibitem{swabianManual}
{Swabian Instruments}, \emph{Time Tagger User Manual}, 2023, \url{https://www.swabianinstruments.com/static/downloads/TimeTagger\_User \_Manual.pdf}.

\bibitem{photon_detect_with_tt}
M.~Wahl, T.~R{\"o}hlicke, S.~Kulisch, S.~Rohilla, B.~Kr{\"a}mer, and A.~C. Hocke, ``Photon arrival time tagging with many channels, sub-nanosecond deadtime, very high throughput, and fiber optic remote synchronization,'' \emph{Review of Scientific Instruments}, vol.~91, no.~1, 2020.

\bibitem{rahmouni2024entangled}
A.~Rahmouni, R.~Wang, J.~Li, X.~Tang, T.~Gerrits, O.~Slattery, Q.~Li, and L.~Ma, ``Entangled photon pair generation in an integrated sic platform,'' \emph{Light: Science \& Applications}, vol.~13, no.~1, p. 110, 2024.

\bibitem{swabianSeries}
{Swabian Instruments}, \emph{Time Tagger Series Datasheet}, Swabian Instruments, 2023, \url{https://www.swabianinstruments.com/static/downloads/TimeTaggerSerie s.pdf}.

\bibitem{picoharp330}
{PicoQuant GmbH}, \emph{PicoHarp 330: Time-Tagged Time-Resolved Data Acquisition Module}, PicoQuant, 2023, \url{https://www.picoquant.com/images/uploads/downloads/datasheet\_picoha rp330.pdf}.

\bibitem{rossetta2021brighteyes}
A.~Rossetta, E.~Slenders, M.~Donato, E.~Perego, F.~Diotalevi, L.~Lanzan{\'o}, S.~Koho, G.~Tortarolo, M.~Crepaldi, and G.~Vicidomini, ``The brighteyes-ttm: an open-source time-tagging module for single-photon microscopy,'' \emph{BioRxiv}, pp. 2021--10, 2021.

\bibitem{ho2009clock}
C.~Ho, A.~Lamas-Linares, and C.~Kurtsiefer, ``Clock synchronization by remote detection of correlated photon pairs,'' \emph{New Journal of Physics}, vol.~11, no.~4, p. 045011, 2009.

\bibitem{enzer2016drifts}
D.~G. Enzer, W.~A. Diener, D.~W. Murphy, S.~R. Rao, and R.~L. Tjoelker, ``Drifts and environmental disturbances in atomic clock subsystems: Quantifying local oscillator, control loop, and ion resonance interactions,'' \emph{IEEE Transactions on Ultrasonics, Ferroelectrics, and Frequency Control}, vol.~64, no.~3, pp. 623--633, 2016.

\bibitem{abane2023data}
A.~Abane, A.~Battou, A.~Amlou, and T.~Zhang, ``A data collection platform for network management,'' in \emph{2023 20th ACS/IEEE International Conference on Computer Systems and Applications (AICCSA)}.\hskip 1em plus 0.5em minus 0.4em\relax IEEE, 2023, pp. 1--7.

\end{thebibliography}
\end{document}